\documentclass[10pt]{article}

\usepackage{graphicx,amsmath,amssymb}
\usepackage{multirow}
\usepackage{color}
\begin{document}

\newtheorem{theorem}{Theorem}[section]
\newtheorem{lemma}[theorem]{Lemma}
\newtheorem{corollary}[theorem]{Corollary}
\newtheorem{proposition}[theorem]{Proposition}
\newcommand{\blackslug}{\penalty 1000\hbox{
    \vrule height 8pt width .4pt\hskip -.4pt
    \vbox{\hrule width 8pt height .4pt\vskip -.4pt
          \vskip 8pt
      \vskip -.4pt\hrule width 8pt height .4pt}
    \hskip -3.9pt
    \vrule height 8pt width .4pt}}
\newcommand{\proofend}{\quad\blackslug}
\newenvironment{proof}{$\;$\newline \noindent {\sc Proof.}$\;\;\;$\rm}{\qed}
\newcommand{\qed}{\hspace*{\fill}\blackslug}
\newtheorem{definition}{Definition}
\def\boxit#1{\vbox{\hrule\hbox{\vrule\kern4pt
  \vbox{\kern1pt#1\kern1pt}
\kern2pt\vrule}\hrule}}
\newcommand{\sm}{\setminus}
\newcommand{\skewregular}{$V_1$-cubic}
\addtolength{\baselineskip}{+0.3mm}

\def\boxit#1{\vbox{\hrule\hbox{\vrule\kern4pt
  \vbox{\kern1pt#1\kern1pt}
\kern2pt\vrule}\hrule}}

\title{\vspace*{-10mm} \bf On Feedback Vertex Set:\\
 New Measure and New Structures\footnote{Supported in part by the US National
 Science Foundation under grants CCF-0830455 and CCF-0917288.}}

\author{ {\sc Yixin Cao}\thanks{Institute for Computer Science and
    Control, Hungarian Academy of Sciences, (MTA SZTAKI,) Budapest,
    Hungary.  Supported in part by the European Research Council (ERC)
    grant 280152 and the Hungarian Scientific Research Fund (OTKA)
    grant NK105645.}
  \and
  {\sc Jianer Chen}\thanks{ Department of Computer Science and
    Engineering, Texas A\&M University, College Station, TX
    77843-3112, USA, {\tt chen@cse.tamu.edu}}
 \and
 {\sc Yang Liu}\thanks{Supercomputing Facility, Texas A\&M University,
   College Station, TX 77845, USA, {\tt yangliu@tamu.edu}} }

\date{}
\maketitle

\begin{abstract}
  We present a new parameterized algorithm for the {feedback vertex set} problem ({\sc fvs}) on undirected graphs. We approach the problem by considering a variation of it, the {disjoint feedback vertex set} problem ({\sc disjoint-fvs}), which finds a feedback vertex set of size $k$ that has no overlap with a given feedback vertex set $F$ of the graph $G$. We develop an improved kernelization algorithm for {\sc disjoint-fvs} and show that {\sc disjoint-fvs} can be solved in polynomial time when all vertices in $G \setminus F$ have degrees upper bounded by three. We then propose a new branch-and-search process on {\sc disjoint-fvs}, and introduce a new branch-and-search measure.  The process effectively reduces a given graph to a graph on which {\sc disjoint-fvs} becomes polynomial-time solvable, and the new measure more accurately evaluates the efficiency of the process. These algorithmic and combinatorial studies enable us to develop an $O^*(3.83^k)$-time parameterized algorithm for the general {\sc fvs} problem, improving all previous algorithms for the problem.
\end{abstract}

\section{Introduction}

All graphs in our discussion are undirected and simple, i.e., they
contain neither self-loops nor multiple edges.  A \emph{feedback
  vertex set} (FVS) $F$ in a graph $G$ is a set of vertices in $G$
whose removal results in an acyclic graph. The problem of finding a
minimum feedback vertex set in a graph is one of the classical
NP-complete problems \cite{GJ79}. It has been intensively studied for
several decades.  The problem is known to be solvable in time
$O(1.7548^n)$ for a graph of $n$ vertices \cite{fomin-08-exact-fvs},
and admit a polynomial-time approximation algorithm of ratio 2
\cite{bafna-99-approximate-fvs,becker-96-approximate-fvs}.

An important application of the feedback vertex set problem is
Bayesian inference in artificial intelligence
\cite{becker-00-randomized-fvs,becker-96-approximate-fvs}, where the
{\it size} $k$ of a minimum FVS $F$ (i.e., the number of vertices in
$F$) of a graph can be expected to be fairly small. This motivated the
study of the parameterized version of the problem, which we will name
{\sc fvs}: given a graph $G$ and a parameter $k$, either construct a
FVS of size bounded by $k$ in $G$ or report no such a FVS
exists. Parameterized algorithms for {\sc fvs} have been extensively
studied that find a FVS of size $k$ in a graph of $n$ vertices in time
$f(k)n^{O(1)}$ for a fixed function $f$ (thus, the algorithms become
practically efficient when the value $k$ is small). The existence of
such an algorithm for {\sc fvs} is implied in
\cite{fellows-88-nonconstructive-polynomial}.  The first group of
constructive algorithms for this problem was given by Downey and
Fellows \cite{DF92} and by Bodlaender
\cite{bodlaender-94-disjoint-cycles}.  Since then a chain of
improvements has been obtained (see
Figure~\ref{fig:table}).\footnote{Following the recent convention in
  the research in exact and parameterized algorithms, we will denote
  by $O^*(f(k))$ the complexity $O(f(k)n^{O(1)})$ for a
  super-polynomial function $f$.}

\begin{figure*}[ht]
\label{fig:table}
\begin{center}
\begin{tabular}{l l r}
  \hline
  {\bf Authors}  & {\bf Complexity} & {\bf Year} \\
  \hline
  Downey and Fellows \cite{DF92} & $O^*((2k+1)^k)$ & 1992 \\
  Bodlaender\cite{bodlaender-94-disjoint-cycles} & $O^*(17 (k^4)!)$ & 1994 \\
  Raman et al.\cite{raman-02-fvs} & $O^*(\max\{12^k,(4\log k)^k\})$ & 2002 \\
  Kanj et al.\cite{kanj-04-fvs} & $O^*((2\log k + 2 \log \log k + 18)^k)$ & 2004 \\
  Raman et al.\cite{raman-06-fvs} & $O^*((12 \log k/\log \log k  + 6)^k)$ & 2006 \\
  Guo et al.\cite{guo-06-fvs} & $O^*(37.7^k)$ & 2006 \\
  Dehne et al.\cite{dehne-05-fvs} & $O^*(10.6^k)$ & 2007 \\
  Chen et al.\cite{chen-08-ufvs} & $O^*(5^k)$ & 2008 \\
  This paper & $O^*(3.83^k)$ & \\
  \hline
\end{tabular}
\end{center}

\vspace*{-2mm}

\caption{The history of parameterized algorithms for {\sc fvs}.}
\end{figure*}

All algorithms summarized in Figure~\ref{fig:table} are {\it
  deterministic}. There is also an active research line on randomized
parameterized algorithms for {\sc fvs}, based on very different
algorithmic techniques. A randomized algorithm of time $O^*(4^k)$ for
{\sc fvs} has been known for more than a decade
\cite{becker-00-randomized-fvs}. More recently, Cygan {et
  al.}~\cite{cygan-11-connectivity-treewidth} developed an improved
randomized algorithm of time $O^*(3^k)$.  As pointed out in
\cite{cygan-11-connectivity-treewidth}, however, the techniques
employed by this randomized algorithm do not seem to be easily
de-randomized.

The main result of the current paper is a deterministic algorithm of time
$O^*(3.83^k)$ for {\sc fvs}.

We give an outline to explain how our algorithm achieves the
improvement over previous algorithms.  As most recent algorithms, our
algorithm is based on the technique of iterative compression
\cite{reed-04-odd-cycle-transversals}, which reduces the {\sc fvs}
problem to a closely related {\sc disjoint feedback vertex set}
problem ({\sc disjoint-fvs}). On an instance $(G, k, F)$, where $F$ is
a FVS in the graph $G$ and $k$ is the parameter, the {\sc
  disjoint-fvs} problem asks whether there is a FVS $F'$ of size $k$
in $G$ such that $F' \cap F = \emptyset$.

The {\sc disjoint-fvs} problem can be solved based on a
branch-and-search process on vertices $w$ in $G \setminus F$, whose
complexity depends on the number of neighbors of $w$ that are in $F$
\cite{chen-08-ufvs}. In particular, the more neighbors $w$ has in $F$,
the more effective the branching on $w$ is. A major step of the
fastest algorithm \cite{chen-08-ufvs}, before our algorithm, is to
show that such a branch-and-search process can always branch on a
vertex in $G \setminus F$ that has at least two neighbors in
$F$. Therefore, in order to further speedup this process, we should
branch only on vertices in $G \setminus F$ that have more than two
neighbors in $F$. For this, however, two issues must be addressed: (1)
during the branch-and-search process, we must be able to continuously
maintain the condition that such vertices always exist; and (2) when
the branch-and-search process cannot be further applied, we must be
able to efficiently solve the problem for the remaining structure.

To address issue (2), we develop a polynomial-time algorithm for the
{\sc disjoint-fvs} problem for instances $(G, k, F)$ in which all
vertices in $G \setminus F$ have degree upper bounded by three.  This
algorithm is based on a nontrivial reduction from {\sc disjoint-fvs}
to a polynomial-time solvable matroid matching problem, the {\sc
  cographic matroid parity} problem
\cite{lovasz-80-matroid-matching}. This result, however, does not give
a direct solution to issue (1): vertices in $G \setminus F$ that have
degree larger than three in $G$ do not necessarily have more than two
neighbors in $F$. To resolve this problem, we observe that there are
always vertices in $G \setminus F$ on which a branching may not be
very effective but will produce structures in $G \setminus F$ that are
favored for the polynomial-time algorithm we developed for addressing
issue (2). To catch this observation, we use the measure-based method
and introduce a new measure to evaluate the effectiveness of our
branch-and-search process more accurately. These new techniques,
combined with the iterative compression method, yield an improved
algorithm for the {\sc fvs} problem.

The main results of this paper are summarized as follows: (i) a new
technique that produces an improved kernelization algorithm for the
{\sc disjoint-fvs} problem, which is based on a branch-and-search
algorithm for the problem. This, to our best knowledge, is the first
time such a technique is used in the literature of kernelization; (ii)
a polynomial-time algorithm that solves a restricted version of the
{\sc disjoint-fvs} problem; (iii) a new branch-and-search process that
effectively reduces an input instance of {\sc disjoint-fvs} to an
instance that is solvable by the algorithm developed in (ii); and (iv)
a new measure that more accurately evaluates the efficiency of the
branch-and-search process in (iii).

\section{{\sc disjoint-fvs} and its kernel}
We start with a formal definition of our first problem.
\begin{quote}
{\sc disjoint-fvs}.\ Given a graph $G = (V, E)$, a FVS $F$ in $G$, and a
parameter $k$, either construct a FVS $F'$ of size $k$ in $G$ such that
$F' \cap F = \emptyset$, or report that no such a FVS $F'$ exists.
\end{quote}

The {\sc disjoint-fvs} problem was motivated by the iterative
compression method \cite{reed-04-odd-cycle-transversals} that has
become a standard framework for the development of parameterized
algorithms for the {\sc fvs} problem. In this method, a critical step
is to construct a solution to an instance $(G, F, k)$ of the {\sc
  disjoint-fvs} problem in which the FVS $F$ satisfies $|F| = k+1$
(see, e.g., \cite{chen-08-ufvs}). However, in the following
discussion, we consider a slightly more generalized version in which
we do not require $|F| = k+1$.

Let $V_1 = V \setminus F$. Since $F$ is a FVS, the subgraph induced
by $V_1$ is a forest. Moreover, if the subgraph induced by $F$
is not a forest, then it is impossible to have a FVS $F'$ in $G$
such that $F' \cap F = \emptyset$. Therefore, an instance of
{\sc disjoint-fvs} can be written as $(G; V_1, V_2; k)$, and
consists of a partition $(V_1, V_2)$ of the vertex set of the graph
$G$ and a parameter $k$ such that both $V_1$ and $V_2$ induce
forests (where $V_2 = F$). We will call a FVS entirely contained in
$V_1$ a {\it $V_1$-FVS}. Thus, the instance $(G; V_1, V_2; k)$ of
{\sc disjoint-fvs} is looking for a $V_1$-FVS of size $k$ in the
graph $G$.

For a subgraph $G'$ of $G$ and a vertex $v$ in $G'$, we will denote by
$d_{G'}(v)$ the degree of the vertex $v$ in $G'$.  Thus, $d_G(v)$ is
the degree of the vertex $v$ in the original graph $G$, and
$d_{G[V_1]}(v)$ for a vertex $v \in V_1$ is the degree of the vertex
$v$ in the induced subgraph $G[V_1]$.

Given an instance $(G; V_1, V_2; k)$ of {\sc disjoint-fvs}, we apply
the following two simple rules:
\begin{quote}
{\bf Rule 1.} \ Remove all vertices $v$ with $d_G(v) \leq 1$;

{\bf Rule 2.} \  For a  vertex $v$ in $V_1$ with $d_G(v) = 2$,

$\bullet$ \ if the two neighbors of $v$ are in the same component
 of $G[V_2]$, then include $v$ into the \\
\hspace*{2.5mm} objective $V_1$-FVS, $G = G - v$, and
$k=k-1$;

$\bullet$ \ else either (2.1) move $v$ from $V_1$ to $V_2$:
     $V_1 = V_1 \setminus \{v\}$, $V_2 = V_2 \cup \{v\}$;
     or (2.2) smoothen $v$: \\
\hspace*{2.5mm} replace $v$ and the two incident edges
     with a new edge connecting the two neighbors of $v$.
\end{quote}

Note that the second case in Rule 2 includes the cases where the two neighbors
of $v$ are both in $V_1$, or both in $V_2$, or one in $V_1$ and one in $V_2$.
In this case, we can pick any of the rules 2.1 and 2.2 and apply it.

The correctness of Rule 1 is trivial: no degree-0 or degree-1
vertices can be contained in any cycle. On the other hand, although
Rule 2 is also easy to verify for the general {\sc fvs} problem
\cite{chen-08-ufvs} (because any cycle containing a degree-2 vertex $v$
must also contain the two neighbors of $v$), it is much less obvious
for the {\sc disjoint-fvs} problem -- the two neighbors of the
degree-2 vertex $v$ may not be in $V_1$ and cannot be included in the
objective $V_1$-FVS. For this, we have the following lemmas.

\begin{lemma}\label{JChen}
For any degree-$2$ vertex $v$ in $V_1$ whose two neighbors are not in the
same component of $G[V_2]$, if $G$ has a $V_1$-FVS of size $k$, then $G$ has a
$V_1$-FVS of size $k$ that does not contain the vertex $v$.

\begin{proof}
Let $F'$ be a $V_1$-FVS of size $k$ that contains $v$.  If one neighbor
$u_1$ of $v$ is in $V_1$, then the set $(F' \setminus \{v\}) \cup
\{u_1\}$ will be a $V_1$-FVS of size bounded by $k$ that does not
contain the vertex $v$.  Thus, we can assume that the two neighbors
$u_1$ and $u_2$ of $v$ are in two different components in
$G[V_2]$. Since $G - F'$ is acyclic, there is either no path or a
unique path in $G - F'$ between $u_1$ and $u_2$. If there is no path
between $u_1$ and $u_2$ in $G - F'$, then adding $v$ to $G - F'$ does
not create any cycle. Therefore, in this case, the set $F' \setminus
\{v\}$ is a $V_1$-FVS of size $k-1$ that does not contain $v$. If
there is a unique path $P$ between $u_1$ and $u_2$ in $G - F'$, then
the path $P$ must contain at least one vertex $w$ in $V_1$ (since
$u_1$ and $u_2$ are in different components in $G[V_2]$). Every cycle
$C$ in $G - (F' \setminus \{v\})$ must contain $v$, thus, also contain
$u_1$ and $u_2$. Therefore, the partial path $C \setminus v$ from
$u_1$ to $u_2$ in $C$ must be the unique path $P$ between $u_1$ and
$u_2$ in $G - F'$, which contains the vertex $w$. This shows that $w$
must be contained in all cycles in $G - (F' \setminus \{v\})$. In
consequence, the set $(F' \setminus \{v\}) \cup \{w\}$ is a $V_1$-FVS
of size bounded by $k$ that does not contain $v$.
\end{proof}
\end{lemma}

\begin{lemma}
\label{lem-ker-2degsafty}
Rule 2 is safe. That is, suppose that Rule 2 applied on $(G; V_1, V_2; k)$
produces $(G'; V_1', V_2'; k')$, then the graph $G'$
has a $V_1'$-FVS of size $k'$ if and only if the graph $G$ has a $V_1$-FVS of size $k$.

\begin{proof}
  If the two neighbors of the degree-$2$ vertex $v$ are contained in
  the same component in $G[V_2]$, then $v$ and some vertices in $V_2$
  form a cycle. Therefore, in order to break this cycle, the vertex
  $v$ must be contained in the objective $V_1$-FVS. This justifies the
  first case for Rule 2.

  If the two neighbors of the degree-$2$ vertex $v$ are not in the
  same component in $G[V_2]$, then $(G'; V_1', V_2'; k')$ is obtained
  by applying either Rule 2.1 or Rule 2.2 on $(G; V_1, V_2; k)$. By
  Lemma~\ref{JChen}, the graph $G$ has a $V_1$-FVS of size $k$ if and
  only if $G$ has a $V_1$-FVS $F_1$ of size $k$ that does not contain
  the vertex $v$. Now it is easy to verify that no matter which of
  Rule 2.1 and Rule 2.2 is applied, we have $k' = k$, and the
  $V_1$-FVS $F_1$ for $G$ becomes a $V_1'$-FVS of size $k$ for the
  graph $G'$. This justifies the second case for Rule 2.
\end{proof}
\end{lemma}

Note that the second case of Rule 2 cannot be applied {\it
simultaneously} on more than one vertex in $V_1$. For example, let
$v_1$ and $v_2$ be two degree-2 vertices in $V_1$ that are both
adjacent to two vertices $u_1$ and $u_2$ in $V_2$. Then it is
obvious that we cannot move both $v_1$ and $v_2$ to $V_2$. In fact,
if we first apply the second case of Rule 2 on $v_1$, then the first
case of Rule 2 will become applicable on the vertex $v_2$.

\begin{definition}
  An instance $(G; V_1, V_2; k)$ of {\sc disjoint fvs} is {\it
    $V_1$-irreducible} if none of the Rules 1-2 can be applied on
  vertices in the set $V_1$, or, equivalently, if all vertices in
  $V_1$ have degree larger than $2$. An instance $(G; V_1, V_2; k)$ is
  {\it nearly $V_1$-irreducible} if in the set $V_1$ there is at most
  one vertex of degree $2$ and all other vertices in $V_1$ are of
  degree larger than $2$.
\end{definition}

For an instance $(G; V_1, V_2; k)$ that is $V_1$-irreducible or nearly
$V_1$-irreducible, in case there is no ambiguity, we will simply say that
the graph $G$ is $V_1$-irreducible or nearly $V_1$-irreducible,
respectively. In the following, we show that a nearly $V_1$-irreducible
instance is necessarily small.

We start with a simple branch-and-search algorithm for nearly
$V_1$-irreducible instances of {\sc disjoint-fvs}, as given in
Figure~\ref{kernel}, which is similar to the one presented in
\cite{chen-08-ufvs}, but gives degree-$2$ vertices a higher priority
when selecting a vertex for branching. The basic step of the algorithm
is to pick a vertex $v$ in $V_1$ and branch on either including or
excluding $v$ in the objective $V_1$-FVS $F$.  Note that in certain
situations, the algorithm directly takes one of the two actions in the
branching (see the footnotes in the algorithm).

\begin{figure}[h]
\setbox4=\vbox{\hsize28pc \noindent\strut
\begin{quote}
\vspace*{-5mm} \small
{\bf Algorithm FindFVS}\\
{\sc input}: a nearly $V_1$-irreducible instance $(G; V_1, V_2; k)$ of
      {\sc disjoint-fvs}.\\
{\sc output}: a $V_1$-FVS $F$ of size $\leq k$ in $G$, or report
      that no such $V_1$-FVS exists.

1. \hspace*{3mm} $F = \emptyset$;\\
2. \hspace*{3mm} {\bf while} $|V_1| > 0$ and $k \geq 0$ {\bf do} \\
3. \hspace*{6mm} {\bf if} there are vertices in $V_1$ that have degree $2$ in $G$ \\
\textcolor{white}{3.} \hspace*{6mm} {\bf then} let $v$ be a vertex in $V_1$ that has degree 2 in $G$ \\
\textcolor{white}{3.} \hspace*{6mm} {\bf else} let $v$ be a vertex in $V_1$ that has degree $\leq 1$ in the induced subgraph $G[V_1]$ \\
4. \hspace*{6mm} {\bf branching} \\
5. \hspace*{10mm} {\bf case 1:} $\setminus\setminus$ $v$ is in the
     objective $V_1$-FVS $F$. \\
6. \hspace*{13.5mm} add $v$ to $F$ and delete $v$ from $G$; \ $k = k-1$;$^\dag$ \\
7. \hspace*{10mm} {\bf case 2:} $\setminus\setminus$ $v$ is not in the
     objective $V_1$-FVS $F$. \\
8. \hspace*{13.5mm}  move $v$ from $V_1$ to $V_2$;$^\ddag$ \\
9. \hspace*{6mm} {\bf if} $|V_1| = 0$ {\bf then} return $F$
     {\bf else} return ``no $V_1$-FVS of size $\leq k$''. \\ \\
$^\dag$ this action will not be taken if $d_G(v) = 2$ and the two neighbors of $v$ are not in
   the same\\ \hspace*{2mm} component of $G[V_2]$. \\
$^\ddag$ this action will not be taken if two neighbors of $v$ are in the same
   component of $G[V_2]$.
\end{quote} \vspace*{-6mm} \strut} $$\boxit{\box4}$$
\vspace*{-9mm}
\caption{A simple branch-and-search algorithm for {\sc disjoint-fvs}}
\label{kernel}
\end{figure}

We will use algorithm FindFVS to count the number of vertices in the set $V_1$.
Note that Rules 1-2 are {\it not} applied during the process of the algorithm.
Initially, the input graph is $V_1$-irreducible. Thus, the selection of the
vertex $v$ in step 3 is always possible. In later steps, the selection of the
vertex $v$ in step 3 can be argued with the following lemma.

\begin{lemma} \label{chennew101}
Each execution of steps 4-8 of algorithm {FindFVS} results
in a nearly $V_1$-irreducible instance.

\begin{proof}
  Since the input instance is nearly $V_1$-irreducible, it suffices to
  prove that on a nearly $V_1$-irreducible instance, the execution of
  steps 4-8 of the algorithm produces a nearly $V_1$-irreducible
  instance. Let $(G; V_1, V_2; k)$ be a nearly $V_1$-irreducible
  instance of {\sc disjoint-fvs} before the execution of steps 4-8 of
  the algorithm, and let $v$ be the vertex in $V_1$ selected by steps
  3 of the algorithm.

  Steps 4-8 either deletes the vertex $v$ from the graph (case 1, steps 5-6) or
  moves $v$ from set $V_1$ to set $V_2$ (case 2, steps 7-8). Moving $v$ from
  $V_1$ to $V_2$ does not change the degree of any vertex remaining in $V_1$.
  Therefore, steps 7-8 keep the resulting instance nearly $V_1$-irreducible.

  Now consider steps 5-6 in the algorithm that delete
  the vertex $v$ from the graph. If $d_G(v) = 2$ and the two neighbors
  of $v$ are in the same component of $G[V_2]$, or if $v$
  has degree $0$ in $G[V_1]$, then
  deleting $v$ does not affect the degree of any vertex remaining in
  $V_1$. Therefore, in these cases steps 5-6 in
  the algorithm produce a nearly $V_1$-irreducible instance. Note
  that by the first footnote in the algorithm, if $d_G(v) = 2$ and the
  two neighbors of $v$ are not in the same component of
  $G[V_2]$, then steps 5-6 of the algorithm will not be
  taken. Therefore, the only remaining case we need to examine is that
  $d_G(v) \geq 3$ and $d_{G[V_1]}(v) \geq 1$. By step 3 of the
  algorithm, in this case, we must have $d_{G[V_1]} = 1$. Let $w$ be
  the unique neighbor of $v$ in $G[V_1]$. By the way we picked the
  vertex $v$ and by our assumption $d_G(v) \geq 3$, no vertex in $V_1$
  has degree $2$ in $G$.  In particular, $d_G(w) \geq 3$. Therefore,
  deleting the vertex $v$ can result in at most one degree-$2$ vertex
  in $V_1$ (i.e., $w$) and will keep all other vertices in $V_1$ with
  degree at least $3$. Thus, in this case steps 5-6 of
  the algorithm again produce a nearly $V_1$-irreducible instance.

  Finally, note that the second footnote in the algorithm ensures that steps 7-8
  will not be taken if the two neighbors of $v$ are in the same component in
  $G[V_2]$. Moreover, steps 4-8 keep $G$ a simple graph since
  they never smoothen vertices. These ensure that steps 4-8 produce a valid
  instance of {\sc disjoint-fvs}.
\end{proof}
\end{lemma}

We make some comments on the algorithm FindFVS. First of all, if there
is no vertex in $V_1$ that has degree $2$ in $G$, then the third line
in step 3 must be able to find a vertex of degree $\leq 1$ in the
subgraph $G[V_1]$ since $V_1$ induces a forest.  Now consider the
correctness of the actions taken in branching steps 4-8. By the
footnotes given in the algorithm FindFVS, if the selected vertex $v$
has degree $2$ in $G$, then no branching is taken and only one of the
cases 1-2 is executed: (1) if both neighbors of $v$ are in the same
component of $G[V_2]$, then only steps 5-6 for case 1 are executed,
i.e., the vertex $v$ is directly included in the objective FVS $F$;
and (2) if the two neighbors of $v$ are not in the same component of
$G[V_2]$, then only steps 7-8 for case 2 are executed, i.e., the
vertex $v$ is moved from $V_1$ to $V_2$. The correctness of the
algorithm FindFVS for these cases is guaranteed by
Lemma~\ref{lem-ker-2degsafty}, which ensures the safeness of Rule
2. When the selected vertex $v$ has a degree different from $2$, then
the branching steps 4-8 are exhaustive and consider both the cases
where $v$ is and is not in the objective FVS. Thus, one of these
actions must be correct. Therefore, if the graph $G$ has a $V_1$-FVS
of size $k$, then one of the computational paths in the search tree
corresponding to the algorithm FindFVS must correctly find such a
$V_1$-FVS.

\begin{theorem}
Let $(G; V_1, V_2; k)$ be a nearly $V_1$-irreducible instance of the
{\sc disjoint-fvs} problem, and let $\tau_1$ and $\tau_2$ be the number
of components in the induced subgraphs $G[V_1]$ and $G[V_2]$,
respectively. Let $\delta_2$ be the number of vertices in $V_1$ that
have degree $2$ in $G$. If $|V_1| > \delta_2 + 2k + \tau_2 - \tau_1 - 1$,
then there is no $V_1$-FVS of size bounded by $k$ in the graph $G$.

\begin{proof}
  We prove the theorem by induction on the number $|V_1|$ of vertices
  in the set $V_1$. For $|V_1| = 1$, we have $\tau_1 = 1$, and the
  condition $|V_1| > \delta_2 + 2k + \tau_2 - \tau_1 - 1$ implies
  $\delta_2 + 2k + \tau_2 \leq 2$.  Let $w$ be the unique vertex in
  $V_1$.  If $\tau_2 = 0$, then the vertex $w$ in $V_1$ would have
  degree $0$ in $G$ (note that by our assumption, $G$ is a simple
  graph), contradicting the assumption that the graph $G$ is nearly
  $V_1$-irreducible.  Thus, we must have $1 \leq \tau_2 \leq 2$, which
  implies $k = 0$. If $\tau_2 = 1$, then since the vertex $w$ in $V_1$
  has degree at least $2$, two neighbors of $w$ must be in the same
  (and unique) component of $G[V_2]$. If $\tau_2 = 2$, then from
  $\delta_2 + 2k + \tau_2 \leq 2$ we have $\delta_2 = 0$, and the
  vertex $w$ has degree at least $3$, which implies again that at
  least two neighbors of $w$ are in the same component of
  $G[V_2]$. Thus, for both cases of $\tau_2 = 1$ and $\tau_2 = 2$, the
  vertex $w$ in $V_1$ must be included in every $V_1$-FVS for $G$,
  which concludes that no $V_1$-FVS of $G$ can have size bounded by $k
  = 0$.  This verifies the theorem for the case $|V_1| = 1$.

  Now consider the general case of $|V_1| > 1$.  Let $(G; V_1, V_2;
  k)$ be a nearly $V_1$-irreducible instance of {\sc disjoint-fvs} and
  suppose that the graph $G$ has a $V_1$-FVS of size bounded by
  $k$. Since the algorithm {FindFVS} solves {\sc disjoint-fvs}
  correctly, there is a computational path $\cal P$ of the algorithm
  that returns a $V_1$-FVS $F$ with $|F| \leq k$. We consider how the
  path $\cal P$ changes the values of an instance when it executes
  (correctly) the action of one of the cases in steps 4-8 in the
  algorithm. Let $|V_1|$, $\delta_2$, $k$, $\tau_1$, and $\tau_2$ be
  the values before the execution of steps 4-8, and let $|V_1'|$,
  $\delta_2'$, $k'$, $\tau_1'$, and $\tau_2'$ be the corresponding
  values after the execution of steps 4-8. The relations between these
  values are summarized in Figure~\ref{chenfig111}, where many are
  obvious. We give below explanations for some less obvious ones in
  the figure.

  We first consider the case where the computational path $\cal P$
  takes the action of case 2 in the algorithm, i.e., moving the vertex
  $v$ from set $V_1$ to set $V_2$. See Table I in
  Figure~\ref{chenfig111}.

  If $d_G(v) = 2$ and both neighbors $w_1$ and $w_2$ of $v$ are in the
  set $V_2$ (see the 3rd line in Table I in Figure~\ref{chenfig111}),
  then by the second footnote in the algorithm, $w_1$ and $w_2$ must
  belong to two different components of $G[V_2]$.  Therefore, moving
  $v$ from $V_1$ to $V_2$ must decrease $\tau_1$ by $1$ (because $v$
  by itself makes a component in $G[V_1]$) and merge the two
  components of $G[V_2]$ into one (i.e., $\tau_2' = \tau_2 - 1$).

  If $d_G(v) \geq 3$ and $v$ has no neighbor in $V_1$ (see the 5th
  line in Table I in Figure~\ref{chenfig111}), then all neighbors of
  $v$ (there are at least $3$) are in $V_2$. Moreover, by the second
  footnote in the algorithm, no two neighbors of $v$ are in the same
  component of $G[V_2]$.  Therefore, moving $v$ from $V_1$ to $V_2$
  decreases the value $\tau_1$ by $1$ (i.e., $\tau_1' = \tau_1 - 1$)
  and merges at least three components of $G[V_2]$ into one (i.e.,
  $\tau_2' \leq \tau_2 - 2$).

  If $d_G(v) \geq 3$ and $N(v) \cap V_1 \neq \emptyset$, then by step
  3 of the algorithm, $v$ has exactly one neighbor in $V_1$ and at
  least two neighbors in $V_2$.  Therefore, if $v$ is moved from $V_1$
  to $V_2$ (see the 6th line in Table I in Figure~\ref{chenfig111}),
  then the value $\tau_1$ is unchanged (i.e., $\tau_1' = \tau_1$), and
  again by the second footnote in the algorithm, the value $\tau_2$ is
  decreased by at least $1$ (i.e., $\tau_2' \leq \tau_2 - 1$).

  Now consider the case where the computational path $\cal P$ takes
  the action of case 1 in the algorithm, i.e., deleting the vertex $v$
  from the graph $G$. See Table II in Figure~\ref{chenfig111}.  First
  note that by the first footnote in the algorithm, if $v$ has degree
  $2$ and if the two neighbors of $v$ do not belong to the same
  component of $G[V_2]$, then the action of case 1 in the algorithm is
  not taken. In particular, the action of case 1 in the algorithm is
  not applicable under the conditions of the 2nd line and the 4th line
  in Table II in Figure~\ref{chenfig111}.

  If $d_G(v) \geq 3$ and if $v$ has no neighbors in $V_1$ (see the 5th
  line in Table II in Figure~\ref{chenfig111}), then deleting $v$ does
  not change the number of degree-$2$ vertices in $V_1$ (i.e.,
  $\delta_2' = \delta_2 = 0$) but decreases the value $\tau_1$ by $1$
  (i.e., $\tau_1' = \tau_1 -1$, because $v$ by itself makes a
  component in $G[V_1]$).

  Finally, if $d_G(v) \geq 3$ and $N(v) \cap V_1 \neq \emptyset$ (see
  the 6th line in Table II in Figure~\ref{chenfig111}), then by the
  way we picked the vertex $v$, we must have $|N(v) \cup V_1| = 1$.
  Let $w$ be the unique neighbor of $v$ in $V_1$. Then, deleting $v$
  may create at most one degree-$2$ vertex (i.e., $w$) in the set
  $V_1$ (i.e., $\delta_2' \leq \delta_2 + 1$), while not changing the
  values of $\tau_1$ and $\tau_2$.

  This verifies all relations in Tables I and II in
  Figure~\ref{chenfig111}.

\begin{figure}
\begin{center}

Table I. Moving the vertex $v$ from set $V_1$ to set $V_2$

\vspace{1mm}

\begin{tabular}{|c|c|c|c|c|c|c|}
\hline
  degree of $v$  & neighbors of $v$ & $\delta_2'$ & $k'$ & $\tau_1'$ &
     $\tau_2'$ & $V_1'$ \\ \hline \hline
  $d_G(v) = 2$ & $w_1, w_2 \in V_1$ & $\delta_2 - 1$ &
     $k$ & $\tau_1 + 1$ & $\tau_2 + 1$ & $V_1 - \{v\}$ \\ \cline{2-7}
  with neighbors & $w_1, w_2 \in V_2$ & $\delta_2 - 1$ &
     $k$ & $\tau_1 - 1$ & $\tau_2 - 1$ & $V_1 - \{v\}$ \\ \cline{2-7}
  $w_1$ and $w_2$ & $w_1 \in V_1$, $w_2 \in V_2$ & $\delta_2-1$ &
     $k$ & $\tau_1$ & $\tau_2$ & $V_1 - \{v\}$\\ \hline
$d_G(v) \geq 3$ & $|N(v) \cap V_1| = 0$ & $\delta_2$ &
     $k$ & $ \tau_1 - 1$ & $\leq \tau_2 - 2$ & $V_1 - \{v\}$\\ \hline
$d_G(v) \geq 3$ & $|N(v) \cap V_1| = 1$ & $\delta_2$ &
     $k$ & $ \tau_1 $ & $\leq \tau_2 - 1$ & $V_1 - \{v\}$\\ \hline
\end{tabular}

\vspace{5mm}
Table II. Deleting the vertex $v$ in $V_1$ from the graph $G$

\vspace{1mm}

\begin{tabular}{|c|c|c|c|c|c|c|}
\hline
  degree of $v$  & neighbors of $v$ & $\delta_2'$ & $k'$ & $\tau_1'$ &
     $\tau_2'$ & $V_1'$ \\ \hline \hline
  $d_G(v) = 2$ & $w_1, w_2 \in V_1$ &  &
      &  &  &  \\ \cline{2-7}
  with neighbors & $w_1, w_2 \in V_2$ & $\delta_2 - 1$ &
     $k-1$ & $\tau_1 - 1$ & $\tau_2$ & $V_1 - \{v\}$ \\ \cline{2-7}
  $w_1$ and $w_2$ & $w_1 \in V_1$, $w_2 \in V_2$ & &
      & & & \\ \hline
$d_G(v) \geq 3$ & $|N(v) \cap V_1| = 0$ & $\delta_2$ &
     $k-1$ & $ \tau_1 - 1$ & $\tau_2$ & $V_1 - \{v\}$\\ \hline
$d_G(v) \geq 3$ & $|N(v) \cap V_1| = 1$ & $\leq \delta_2 + 1$ &
     $k-1$ & $\tau_1$ & $\tau_2$ & $V_1 - \{v\}$\\ \hline
\end{tabular}
\end{center}
\caption{Results of applying the steps 4-8 of algorithm {FindFVS}
         on vertex $v$}
\label{chenfig111}
\end{figure}

Let $(G'; V_1', V_2'; k')$ be the instance produced by the computational
path $\cal P$ on the nearly $V_1$-irreducible instance
$(G; V_1, V_2; k)$. By our assumption, the graph $G$ has a $V_1$-FVS
of size $k$. Since we also assume that the computational path
$\cal P$ is correct, the graph $G'$ must have a $V_1'$-FVS of size
bounded by $k'$. Since $|V_1'| = |V_1| - 1$ and by Lemma~\ref{chennew101},
the instance $(G'; V_1', V_2'; k')$ is nearly $V_1'$-irreducible, we can
apply the induction on the instance $(G'; V_1', V_2'; k')$,
which gives $|V_1'| \leq \delta_2' + 2k' + \tau_2' - \tau_1' - 1$. This
gives
\[ |V_1| = |V_1'| + 1
   \leq \delta_2' + 2k' + \tau_2' - \tau_1' - 1 + 1. \]
Using this inequality to examine each situation in Figure~\ref{chenfig111},
we can easily verify that the inequality
\[ |V_1| \leq \delta_2 + 2k + \tau_2 - \tau_1 - 1 \]
holds true. Therefore, if $|V_1| > \delta_2 + 2k + \tau_2 - \tau_1 - 1$,
then the graph $G$ has no $V_1$-FVS of size bounded by $k$. This completes
the proof of the theorem.
\end{proof}
\end{theorem}

Since a $V_1$-irreducible instance is also nearly $V_1$-irreducible
in which $\delta_2 = 0$, we get immediately
\begin{corollary}
\label{chencor1}
Let $(G; V_1, V_2; k)$ be a $V_1$-irreducible instance of the
{\sc disjoint-fvs} problem. If $|V_1| > 2k + \tau_2 - \tau_1 - 1$,
then there is no $V_1$-FVS of size bounded by $k$ in the graph $G$.
\end{corollary}

\begin{figure}
\begin{center}
\begin{picture}(200,75)
\put(100,11){\circle{5}} \put(100,28){\circle{5}}
\put(10,28){\circle*{5}} \put(190,28){\circle*{5}}
\put(20,48){\circle*{5}} \put(180,48){\circle*{5}}
\put(40,68){\circle*{5}} \put(160,68){\circle*{5}}
\put(100,14){\line(0,1){12}}
\put(10,28){\line(1,0){87}} \put(190,28){\line(-1,0){87}}
\put(10,28){\line(5,-1){87}} \put(190,28){\line(-5,-1){87}}
\put(10,28){\line(1,2){10}} \put(190,28){\line(-1,2){10}}
\put(20,48){\line(1,1){20}} \put(180,48){\line(-1,1){20}}
\put(98,30){\line(-4,1){80}} \put(102,30){\line(4,1){80}}
\put(98,30){\line(-3,2){60}} \put(102,30){\line(3,2){60}}
\multiput(46,71)(8,0){14}{\line(1,0){4}}
\put(96,0){$w_2$} \put(96,36){$w_1$}
\put(0,23){$v_1$} \put(8,48){$v_2$} \put(27,69){$v_3$}
\put(194,23){$v_{2k-1}$} \put(185,48){$v_{2k-2}$} \put(165,69){$v_{2k-3}$}
\end{picture}
\end{center}
\caption{An example showing the tightness of Corollary~\ref{chencor1}.}
\label{chenfig222}
\end{figure}
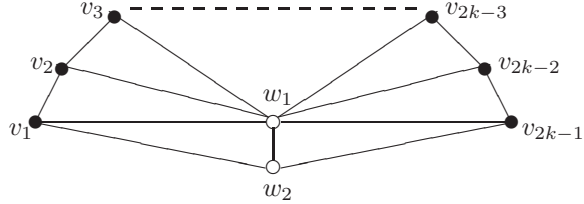

The bound given in Corollary~\ref{chencor1} is in fact {\it tight},
which can be seen as follows.  Consider the graph $G$ in
Figure~\ref{chenfig222}, which consists of $2k+1$ vertices $w_1$,
$w_2$, $v_1$, $v_2$, $\ldots$, $v_{2k-1}$, where $k \geq 2$ is an
arbitrary positive integer. The vertices of $G$ are partitioned into
two sets $V_1 = \{v_1, v_2, \ldots, v_{2k-1}\}$ and $V_2 = \{w_1,
w_2\}$, and $(G; V_1, V_2; k)$ is a $V_1$-irreducible instance of the
{\sc disjoint-fvs} problem. Note that $\tau_1 = \tau_2 = 1$.  We have
$|V_1| = 2k-1 = 2k + \tau_2 - \tau_1 - 1$, while the graph $G$ has a
$V_1$-FVS $F$ of $k$ vertices: $F = \{v_1, v_3, v_5, \ldots,
v_{2k-1}\}$.

A particularly interesting class of instances of the {\sc
  disjoint-fvs} problem was motivated by the iterative compression
method for solving the {\sc fvs} problem, in which each instance $(G;
V_1, V_2; k)$ satisfies an additional condition $|V_2|=k+1$.  Call
this restricted version of {\sc disjoint-fvs} the {\sc
  disjoint-smaller-fvs} problem. For this important version of {\sc
  disjoint-fvs}, we have the following kernelization result.

\begin{theorem}
\label{col:3k-kernel}
The {\sc disjoint-smaller-fvs} problem has a $4k$-vertex kernel: there
is a polynomial-time algorithm that, on an instance $(G; V_1, V_2; k)$
of {\sc disjiont-smaller-fvs}, produces an equivalent instance $(G';
V_1', V_2'; k')$ of {\sc disjiont-smaller-fvs} such that $k' \leq k$
and the graph $G'$ contains at most $4k'$ vertices.

\begin{proof}
  On an instance $(G; V_1, V_2; k)$ of {\sc disjoint-smaller-fvs}, we
  apply Rule 1 and Rule 2 on vertices in $V_1$.  However, for a
  degree-2 vertex $v$ in $V_1$ with neighbors $u_1$ and $u_2$ not in
  the same component of $G[V_2]$, we smoothen $v$ {\it except} in the
  case where $u_1$ is in $V_1$, $u_2$ is in $V_2$, and $[u_1, u_2]$ is
  an edge in $G$. In this case we instead include $u_1$ in the
  objective FVS, and remove both $u_1$ and $v$. This change can be
  justified as follows. By Lemma~\ref{lem-ker-2degsafty}, we can move
  $v$ from $V_1$ to $V_2$, which will make $u_1$ a vertex in $V_1$
  that has two neighbors $v$ and $u_2$ in the same component in
  $G[V_2]$. Thus, $u_1$ can be included directly in the objective FVS,
  and removed. The removal of $u_1$ makes $v$ become a degree-1 vertex
  so can also be removed.

The reason for this change is that we want to keep $G$ a simple graph
without changing the vertex set $V_2$.  Smoothening a degree-2 vertex
$v$ in $V_1$ with neighbors $u_1$ and $u_2$ such that $[u_1, u_2]$ is
an edge will create multiple edges. Note that in this case, (1) $u_1$
and $u_2$ cannot be both in $V_1$ since $V_1$ induces a forest; and
(2) $u_1$ and $u_2$ cannot be both in $V_2$ because otherwise, $v$
would have two neighbors in the same component of $G[V_2]$ and $v$
would be included in the objective FVS. Thus, the only possibility
that this may happen is that one of $u_1$ and $u_2$ is in $V_1$ and
the other is in $V_2$. Thus, the process in the previous paragraph
avoids creating multiple edges, keeps the graph $G$ a simple graph,
and keep the vertex set $V_2$ unchanged (although it may add edges
between vertices in $V_2$ when smoothening degree-2 vertices in
$V_1$).

We repeat this process until it is no longer applicable. Let $(G'';
V_1'', V_2''; k'')$ be the resulting instance. By
Lemma~\ref{lem-ker-2degsafty} and the above discussion, $(G''; V_1'',
V_2''; k'')$ is a YES-instance of {\sc disjoint-fvs} if and only if
$(G; V_1, V_2; k)$ is a YES-instance of {\sc disjoint-small-fvs}.
Moreover, $k'' \leq k$, $V_2'' = V_2$, and all vertices in $V_1''$
have degree at least $3$ in $G''$. Thus, $(G''; V_1'', V_2''; k'')$ is
$V_1''$-irreducible. By Corollary~\ref{chencor1}, we can assume
$|V_1''| \leq 2k'' + \tau_2'' - \tau_1'' - 1$, where $\tau_1''$ and
$\tau_2''$ are the number of components in $G''[V_1'']$ and
$G''[V_2'']$, respectively, for which we have $\tau_2'' \leq |V_2''| =
|V_2| = k+1$ and $\tau_1'' \geq 1$. Thus, the total number $|G''|$ of
vertices in the graph $G''$ is $|V_1''| + |V_2''| \leq (2k'' + (k+1) -
2) + (k+1) = 2(k''+k)$.

However, $(G''; V_1'', V_2''; k'')$ may not be an instance of {\sc
  disjoint-smaller-fvs} because we may have $|V_2''| = |V_2| = k+1 >
k''+1$. If this is the case, let $h = k - k''$, and we add a disjoint
simple path $P_{2h} = (w_1, \ldots w_{2h})$ of $2h$ vertices to $G''$
and let these $2h$ vertices be adjacent to a fixed vertex $u$ in
$V_2''$.  Let the new graph be $G'$, with the vertex partition $(V_1',
V_2')$, where $V_1' = V_1'' \cup \{w_1, \ldots, w_{2h}\}$ and $V_2' =
V_2''$. Now consider the instance $(G'; V_1', V_2'; k')$ of {\sc
  disjoint-fvs}, where $k' = k$. It is easy to verify that the graph
$G'$ has a $V_1'$-FVS of $k' = k$ vertices if and only if the graph
$G''$ has a $V_1''$-FVS of $k'-h = k - h = k''$ vertices. Moreover,
since $|V_2'| = |V_1''| = |V_2| = k+1 = k'+1$, $(G'; V_1', V_2'; k')$
is a valid instance for {\sc disjoint-smaller-fvs}. Therefore, $(G';
V_1', V_2'; k')$ is a YES-instance of {\sc disjoint-smaller-fvs} if
and only if $(G; V_1, V_2; k)$ is a YES-instance of {\sc
  disjoint-smaller-fvs}: this holds true because both of these
conditions are equivalent to the condition that $(G''; V_1'', V_2'';
k'')$ is a YES-instance of {\sc disjoint-fvs}. Finally, the number of
vertices in $G'$ is equal to $|G''| + 2h \leq 2(k''+k) + 2(k-k'') = 4k
= 4k'$.
\end{proof}
\end{theorem}

Finally, we remark that this kernelization result was obtained based
on the branch-and-search algorithm {FindFVS} for the problem, instead
of on an analysis of the resulting structure after applying reduction
rules. This technique, to our best knowledge, had not been used in the
literature of kernelization.

\section{A polynomial-time solvable case for {\sc disjoint-fvs}}

In this section we consider a special class of instances for {\sc
  disjoint-fvs}. This approach is closely related to the classical
study on graph maximum genus embeddings
\cite{chen-03-max-genus,furst88}. However, the study on graph maximum
genus embeddings that is related to our approach is based on general
spanning trees of a graph, while our approach must be restricted to
only spanning trees that are constrained by the vertex partition
$(V_1, V_2)$ of an instance $(G; V_1, V_2; k)$ of {\sc disjoint-fvs}.
We start with a simple lemma.

\begin{lemma} \label{new11}
Let $G$ be a connected graph and let $H$ be a subgraph of $G$ such that
$H$ is a forest. There is a spanning tree in $G$ that contains the entire
subgraph $H$, and can be constructed in time $O(m \alpha(n))$, where
$\alpha(n)$ is the inverse of Ackermann function.

\begin{proof}
The lemma can be proved based on a process that is similar to the
well-known Kruskal's algorithm for constructing a minimum spanning
tree for a given graph, which runs in time
$O(m \alpha(n))$ if we do not have to sort the edges. Starting from
a structure $G_0$ that initially consists of the forest $H$ and
all vertices in $G$ that are not in $H$, we repeatedly add each of
the remaining edges (in an arbitrary order) to the structure $G_0$
as long as the edge does not create a cycle. The resulting structure
of this process must be a spanning tree that contains the entire
subgraph $H$.
\end{proof}
\end{lemma}

Let $(G; V_1, V_2; k)$ be an instance for {\sc disjoint-fvs}. Since
the induced subgraph $G[V_2]$ is a forest, by Lemma~\ref{new11}, there
is a spanning tree $T$ of the graph $G$ that contains $G[V_2]$. Call a
spanning tree that contains $G[V_2]$ a {\it $G[V_2]$-spanning tree}.

For a graph $H$, denote by $E(H)$ the set of edges in $H$, and for an
edge subset $E'$ in $H$, denote by $H - E'$ the graph $H$ with the
edges in $E'$ removed (the end vertices of these edges are not
removed).

Let $T$ be a $G[V_2]$-spanning tree of the graph $G$. By the construction,
every edge in $G - E(T)$ has at least one end in $V_1$. Two edges in
$G - E(T)$ are {\it $V_1$-adjacent} if they have a common end in
$V_1$. A {\it $V_1$-adjacency matching} in $G - E(T)$ is a partition
of the edges in $G - E(T)$ into groups of one or two edges, called
{\it 1-groups} and {\it 2-groups}, respectively, such that two edges
in the same 2-group are $V_1$-adjacent. A {\it maximum $V_1$-adjacency
  matching} in $G - E(T)$ is a $V_1$-adjacency matching in
$G - E(T)$ that maximizes the number of 2-groups.

\begin{definition}
Let $(G; V_1, V_2; k)$ be an instance of the {\sc disjoint-fvs}
problem. The {\it $V_1$-adjacency matching number $\nu(G, T)$} of a
$G[V_2]$-spanning tree $T$ in $G$ is the number of 2-groups in a maximum
$V_1$-adjacency matching in $G - E(T)$. The {\it $V_1$-adjacency
matching number} $\nu(G)$ of the graph $G$ is the largest $\nu(G, T)$
over all $G[V_2]$-spanning trees $T$ in the graph $G$.
\end{definition}

An instance $(G; V_1, V_2; k)$ of {\sc disjoint-fvs} is {\it
  \skewregular} if every vertex in the set $V_1$ has degree
exactly $3$. Let $f_{V_1}(G)$ be the size of a minimum $V_1$-FVS for
$G$. Let $\beta(G)$ be the {\it Betti number} of $G$ that is the total
number of edges in $G - E(T)$ for any spanning tree $T$ in
$G$. Note that the edge set $G - E(T)$ forms a basis of the {\it
  fundamental cycles} for the graph $G$ such that {\it every} cycle in
$G$ contains at least one edge in $G - E(T)$.  In this sense,
$\beta(G)$ is the number of fundamental cycles in the graph $G$
\cite{furst88}.

\begin{lemma}\label{new12}
  For any \skewregular\ instance $(G; V_1, V_2; k)$ of {\sc
    disjoint-fvs}, we have $f_{V_1}(G) = \beta(G) - \nu(G)$.
  Moreover, a minimum $V_1$-FVS of the graph $G$ can be constructed in
  linear time from a $G[V_2]$-spanning tree whose $V_1$-adjacency matching
  number is $\nu(G)$.

\begin{proof}
  First note that a maximum $V_1$-adjacency matching in $G - E(T)$ for
  a $G[V_2]$-spanning tree $T$ can be constructed in linear time, as
  follows.  Since each vertex in $V_1$ has degree $3$ and $T$ is a
  spanning tree in $G$, each vertex in $G - E(T)$ has degree bounded
  by $2$.  Thus, each component of $G - E(T)$ is either a simple
  (possibly trivial) path or a simple cycle.  Therefore, a maximum
  $V_1$-adjacency matching in $G - E(T)$ can be constructed trivially
  by maximally pairing the edges in every component of $G - E(T)$.

  Let $T$ be a $G[V_2]$-spanning tree such that there is a
  $V_1$-adjacency matching $M$ in $G - E(T)$ that contains $\nu(G)$
  2-groups. Let $U$ be the set of edges that are in the 1-groups in
  $M$. We construct a $V_1$-FVS $F$ as follows: (1) for each edge $e$
  in $U$, arbitrarily pick an end of $e$ that is in $V_1$ and include
  it in $F$; and (2) for each 2-group of two $V_1$-adjacent edges
  $e_1$ and $e_1$ in $M$, pick the vertex in $V_1$ that is a common
  end of $e_1$ and $e_2$ and include it in $F$. Note that every cycle
  in the graph $G$ contains at least one edge in $G - E(T)$, while now
  every edge in $G - E(T)$ has at least one end in $F$. Therefore, $F$
  is a FVS. By the above construction, $F$ is a $V_1$-FVS. The number
  of vertices in $F$ is equal to $|U| + \nu(G)$. Since $|U| = |G -
  E(T)| - 2 \nu(G) = \beta(G) - 2\nu(G)$, we have $|F| = \beta(G) -
  \nu(G)$. This concludes that
 \begin{equation}
   f_{V_1}(G) \leq \beta(G) - \nu(G).  \label{eq0}
 \end{equation}

 Now consider the other direction. Let $F$ be a minimum $V_1$-FVS for
 the graph $G = (V, E)$, i.e., $|F| = f_{V_1}(G)$.  By
 Lemma~\ref{new11}, there is a spanning tree $T$ in $G$ that contains
 the entire subgraph $G - {F}$, which is a forest. We construct a
 $V_1$-adjacency matching in $G - E(T)$ and show that it contains at
 least ($\beta(G) - |F|$) 2-groups. Since $T$ contains $G - {F}$, each
 edge in $G - E(T)$ has at least one end in $F$. Let $E_2$ be the set
 of edges in $G - E(T)$ that have their both ends in $F$, and let
 $E_1$ be the set of edges in $G - E(T)$ that have exactly one end in
 $F$.
 \begin{quote}
 {\bf Claim}. \ Each end of an edge in $E_2$ is shared by exactly one
  edge in $E_1$. In particular, no two edges in $E_2$ share a common
  end.
 \end{quote}

To prove the above claim, first note that since $T$ is a spanning tree
in $G$, each vertex in $F \subseteq V_1$, which has degree $3$ in $G$,
can be incident to at most two edges in $G - E(T) = E_1 \cup E_2$. In
particular, if $u$ is an end of an edge $[u, v]$ in $E_2$ (i.e.,
$u, v \in F$), then there is at most one other edge in $E_1 \cup E_2$
that is incident to $u$. Now assume to the contrary of the claim that
the vertex $u$ is not shared by an edge in $E_1$. Then for the other
two edges $e_1$ and $e_2$ in $G$ that are incident to $u$, either both
$e_1$ and $e_2$ are in $T$ or exactly one of $e_1$ and $e_2$ is in
$E_2$. If both $e_1$ and $e_2$ are in $T$, then every edge in
$G - E(T)$ (including $[u,v]$) has at least one end in
$F \setminus \{u\}$. Similarly, if exactly one $[u, w]$ of the edges
$e_1$ and $e_2$ is in $E_2$, where $w$ is also in $F$, then again every
edge in $G - E(T)$ (including $[u,v]$ and $[u, w]$) has at least one
end in $F \setminus \{u\}$. Thus, in either case, $F \setminus \{u\}$
would make a smaller $V_1$-FVS, contradicting the assumption that $F$
is a minimum $V_1$-FVS. This proves the claim.

Suppose that there are $m_2$ vertices in $F$ that are incident to
two edges in $G - E(T)$. Thus, each of the rest $|F|-m_2$ vertices
in $F$ is incident to at most one edge in $G - E(T)$. By counting
the total number of incidencies between the vertices in $F$ and the edges in
$G - E(T)$, we get
 \[ 2 |E_2| + |E_1| = 2 |E_2| + (\beta(G) - |E_2|) \leq 2 m_2 + (|F| - m_2),\]
or equivalently,
 \begin{equation}
   m_2 - |E_2| \geq \beta(G) - |F|.   \label{eq1}
 \end{equation}

Now we construct a $V_1$-adjacency matching in $G - E(T)$, as
follows. For each edge $e$ in $E_2$, by the above claim, we can make
a 2-group that consists of $e$ and an edge in $E_1$ that shares an
end in $V_1$ with $e$ (note that this grouping will not put an edge in
$E_1$ in two different 2-groups because if the edge $e$ in $E_2$ shares
an end with an edge $e'$ in $E_1$, then $e'$ cannot share an end with
any other edges in $E_2$). Besides the ends of the edges in $E_2$, there
are $m_2 - 2|E_2|$ vertices in $F$ that are incident to two edges in
$E_1$. For each $v$ of these vertices, we make a 2-group that consists
of the two edges in $E_1$ that are incident to $v$. Note that this
construction of 2-groups never uses any edges in $G - E(T)$ more
than once. Therefore, the construction gives $|E_2| + (m_2 - 2|E_2|) =
m_2 - |E_2|$ disjoint 2-groups. We then make each of the rest edges
in $G - E(T)$ a 1-group. This gives a $V_1$-adjacency matching in
$G - E(T)$ that has $m_2 - |E_2|$ 2-groups. By Inequality (\ref{eq1})
and by definition, we have
 \begin{equation}
   \nu(G) \geq \nu(G, T) \geq m_2 - |E_2| \geq \beta(G) - |F| =
     \beta(G) - f_{V_1}(G).       \label{eq2}
 \end{equation}
Combining (\ref{eq0}) and (\ref{eq2}), we conclude with
$f_{V_1}(G)=\beta(G)-\nu(G)$.

The first two paragraphs in this proof also illustrate how to construct in
linear time a minimum $V_1$-FVS from a $G[V_2]$-spanning tree whose
$V_1$-adjacency matching number is $\nu(G)$.
\end{proof}
\end{lemma}

By Lemma~\ref{new12}, in order to construct a minimum $V_1$-FVS for a
\skewregular\ instance $(G; V_1, V_2, k)$ of {\sc disjoint-fvs}, we
only need to construct a $G[V_2]$-spanning tree in the graph $G$ whose
$V_1$-adjacency matching number is $\nu(G)$. The construction of an
unconstrained maximum adjacency matching in terms of general spanning
trees has been considered by Furst et al.~\cite{furst88} in their
study of graph maximum genus embeddings.  We follow a similar
approach, based on cographic matroid parity, to construct a
$G[V_2]$-spanning tree in $G$ whose $V_1$-adjacency matching number is
$\nu(G)$. We start with a quick review on the related concepts in
matroid theory. More detailed discussion on this problem can be found
in \cite{lovasz-80-matroid-matching}.

A {\it matroid} is a pair $(E, \Im)$, where $E$ is a finite set and
$\Im$ is a nonempty collection of subsets of $E$ that contains the empty
set $\emptyset$ and satisfies the
following properties (note that the collection $\Im$ may not be
explicitly given but is defined in terms of certain subset
properties):
\begin{quote}
(1) If $A \in \Im$ and $B \subseteq A$, then $B \in \Im$;

(2) If $A, B \in \Im$ and $|A| > |B|$, then there is an element $a
\in A \setminus B$ such that $B \cup \{a\} \in \Im$.
\end{quote}
The {\it matroid parity} problem is stated as follows: given a
matroid $(E, \Im)$ and a perfect pairing $\{[a_1, \overline{a}_1]$,
$[a_2, \overline{a}_2]$, $\ldots, [a_n, \overline{a}_n]\}$ of the
elements in the set $E$, find a largest subset $M$ in $\Im$ such
that for all $i$, $1 \leq i \leq n$, either both $a_i$ and
$\overline{a}_i$ are in $M$, or neither of $a_i$ and
$\overline{a}_i$ is in $M$.

Each connected graph $G$ is associated with a {\it cographic matroid}
$(E_G, \Im_G)$, where $E_G$ is the edge set of $G$, and an edge set
$S$ is in $\Im_G$ if and only if $G - S$ is connected. It is
well-known that matroid parity problem for cographic matroids can be
solved in polynomial time \cite{lovasz-80-matroid-matching}.  The
fastest known algorithm for cographic matroid parity problem is by
Gabow and Xu \cite{gabow-96-efficient-matroid-intersection},
which runs in time $O(mn\log^6n)$.

In the following, we explain how to reduce our problem to the
cographic matroid parity problem. Let $(G; V_1, V_2; k)$ be a
\skewregular\ instance of the {\sc disjoint-fvs} problem. Without loss
of generality, we make the following assumptions: (1) the graph $G$ is
connected (otherwise, we simply work on each component of $G$); and
(2) for each vertex $v$ in $V_1$, there is at most one edge from $v$
to a component in $G[V_2]$ (otherwise, we can directly include $v$ in
the objective $V_1$-FVS).

Recall that two edges are {\it $V_1$-adjacent} if they share a
common end in $V_1$. For an edge $e$ in $G$, denote by $d_{V_1}(e)$
the number of edges in $G$ that are $V_1$-adjacent to $e$ (note that
an edge can be $V_1$-adjacent to the edge $e$ from either end of
$e$).

We construct a {\it labeled subdivision $G_2$} of the graph $G$ as
follows.
\begin{enumerate}
 \item shrink each component of $G[V_2]$ into a single
  vertex; let the resulting graph be $G_1$;
\vspace*{-2mm}
\item assign each edge in $G_1$ a distinguished label;
\vspace*{-2mm}
\item for each edge labeled $e_0$ in $G_1$, suppose the edges
  $V_1$-adjacent to $e_0$ are labeled by $e_1$, $e_2$, $\ldots$,
  $e_d$ (in arbitrary order), where $d = d_{V_1}(e_0)$; subdivide
  $e_0$ into $d$ {\it segment edges} by inserting $d-1$ degree-2
  vertices in $e_0$, and label the segment edges by $(e_0e_1)$,
  $(e_0e_2)$,  $\ldots$, $(e_0e_d)$. Let the resulting graph be
  $G_2$. The segment edges $(e_0e_1)$, $(e_0e_2)$,  $\ldots$,
  $(e_0e_d)$ in $G_2$ are said to be {\it from} the edge $e_0$ in
  $G_1$.
\end{enumerate}

There are a number of interesting properties for the graphs
constructed above. First, each of the edges in the graph $G_1$
corresponds uniquely to an edge in $G$ that has at least one end in
$V_1$. Thus, without creating any confusion, we will simply say that
the edge is in the graph $G$ or in the graph $G_1$. Second, because
of the assumptions we made on the graph $G$, the graph $G_1$ is a
simple and connected graph. In consequence, the graph $G_2$ is also
a simple and connected graph. Finally, because each edge in $G_1$
corresponds to an edge in $G$ that has at least one end in $V_1$,
and because each vertex in $V_1$ has degree 3, every edge in $G_1$
is subdivided into at least two segment edges in $G_2$.

Now in the labeled subdivision graph $G_2$, pair the segment edge
labeled $(e_0e_i)$ with the segment edge labeled $(e_ie_0)$ for all
segment edges (note that $(e_0e_i)$ is a segment edge from the edge
$e_0$ in $G_1$ and that $(e_ie_0)$ is a segment edge from the edge
$e_i$ in $G_1$). By the above remarks, this is a perfect pairing $\cal
P$ of the edges in $G_2$. Now with this edge pairing $\cal P$ in
$G_2$, and with the cographic matroid $(E_{G_2}, \Im_{G_2})$ for the
graph $G_2$, we call Gabow and Xu's algorithm
\cite{gabow-96-efficient-matroid-intersection} for the cographic
matroid parity problem. The algorithm produces a maximum edge subset
$M$ in $\Im_{G_2}$ that, for each segment edge $(e_0e_i)$ in $G_2$,
either contains both $(e_0e_i)$ and $(e_ie_0)$, or contains neither of
$(e_0e_i)$ and $(e_ie_0)$.

\begin{lemma} \label{chennew}
From the edge subset $M$ in $\Im_{G_2}$ constructed above, a
$G[V_2]$-spanning tree for the graph $G$ with a $V_1$-adjacency matching
number $\nu(G)$ can be constructed in time $O(m \alpha(n))$,
where $n$ and $m$ are the number of vertices and the number of
edges, respectively, of the original graph $G$.

\begin{proof}
  Suppose that the edge subset $M$ consists of the edge pairs $\{
  [(e_1e_1'), (e_1'e_1)], \ldots, [(e_he_h'), (e_h'e_h)]\}$ in
  $G_2$. Since $M \in \Im_{G_2}$, $G_2 - M$ is connected. Thus, for
  each edge $e_i$ in $G_1$, there is at most one segment edge in $M$
  that is from $e_i$.  Therefore, the edge subset $M$ corresponds to
  an edge subset $M'$ of exactly $2h$ edges in $G_1$ (thus exactly
  $2h$ edges in $G$): $M' = \{ e_1, e_1'; \ldots, e_h, e_h'\}$, where
  for $1 \leq i \leq h$, the edges $e_i$ and $e_i'$ are
  $V_1$-adjacent. Since $G_2-M$ is connected, it is easy to verify
  that the graph $G_1-M'$ (thus the graph $G-M'$) is also
  connected. Also note that the graph $G-M'$ contains the induced
  subgraph $G[V_2]$ because no edge in $G_1$ has its both ends in
  $V_2$. Therefore, by Lemma~\ref{new11}, we can construct, in time
  $O(m \alpha(n))$, a $G[V_2]$-spanning tree $T_1$ for the graph $G -
  M'$, which is also a $G[V_2]$-spanning tree for the graph $G$. Now if
  we make each pair $[e_i, e_i']$ a 2-group for $1 \leq i \leq h$, and
  make each of the rest edges in $G - E(T_1)$ a $1$-group, we get a
  $V_1$-adjacency matching with $h$ 2-groups in $G - E(T_1)$.

  To complete the proof of the lemma, we only need to show that $h =
  \nu(G)$. For this, it suffices to show that no $G[V_2]$-spanning tree
  can have a $V_1$-adjacency matching with more than $h$ 2-groups. Let
  $T_2$ be a $G[V_2]$-spanning tree with $q$ 2-groups $[e_1, e_1']$,
  $\ldots$, $[e_q, e_q']$ in $G - E(T_2)$. Since $G - \bigcup_{i=1}^q
  \{e_i, e_i'\}$ entirely contains $T_2$, it is connected. In
  consequence, the graph $G_1 - \bigcup_{i=1}^q \{e_i, e_i'\}$ is also
  connected. From this, it is easy to verify that the graph $G_2 -
  \bigcup_{i=1}^q \{(e_ie_i'), (e_i'e_i)\}$ is also
  connected. Therefore, the edge subset $\{(e_1e_1'), (e_1'e_1);
  \ldots, (e_qe_q'), (e_q'e_q)\}$ is in $\Im_{G_2}$. Now since $M$ is
  the the solution of the matroid parity problem for the cographic
  matroid $(E_{G_2}, \Im_{G_2})$ and since $M$ consists of $h$ edge
  pairs, we must have $h \geq q$. This completes the proof of the
  lemma.
\end{proof}
\end{lemma}

Now we are ready to present our main result in this section, which is
a nontrivial generalization of a result in \cite{ueno-88-cubic-fvs}
(the result in \cite{ueno-88-cubic-fvs} can be viewed as a special
case of Lemma~\ref{new12} in which all vertices in the set $V_2$ have
degree 2).

\begin{theorem}\label{new13}
  There is an $O(n^2 \log^6 n)$-time algorithm that on a
  \skewregular\ instance $(G; V_1, V_2; k)$ of {\sc disjoint-fvs},
  either constructs a $V_1$-FVS of size bounded by $k$, if such a
  $V_1$-FVS exists, or reports correctly that no such a $V_1$-FVS
  exists.

\begin{proof}
  For the \skewregular\ instance $(G; V_1, V_2; k)$ of {\sc
    disjoint-fvs}, we first construct the graph $G_1$ in linear time
  by shrinking each component of $G[V_2]$ into a single vertex.  Note
  that since each vertex in $V_1$ has degree 3, the total number of
  edges in $G_1$ is bounded by $3 |V_1|$. From the graph $G_1$, we
  construct the labeled subdivision graph $G_2$. Again since each
  vertex in $V_1$ has degree 3, each edge in $G_1$ is subdivided into
  at most 4 segment edges in $G_2$. Therefore, the number $n_2$ of
  vertices and the number $m_2$ of edges in $G_2$ are both bounded by
  $O(|V_1|) = O(n)$. From the graph $G_2$, we apply Gabow and
  Xu's algorithm \cite{gabow-96-efficient-matroid-intersection}
  on the cographic matroid $(E_{G_2}, \Im_{G_2})$ that produces the
  edge subset $M$ in $\Im_{G_2}$ in time $O(m_2 n_2 \log^6 n_2) =
  O(n^2 \log^6 n)$.  By Lemma~\ref{chennew}, from the edge subset $M$,
  we can construct in time $O(m \alpha(n))$ a $G[V_2]$-spanning tree $T$
  for the graph $G$ whose $V_1$-adjacency matching number is
  $\nu(G)$. Finally, by Lemma~\ref{new12}, from the
  $G[V_2]$-spanning tree $T$, we can construct a minimum $V_1$-FVS $F$
  in linear time. Now the solution to the \skewregular\ instance $(G;
  V_1, V_2; k)$ of {\sc disjoint-fvs} can be trivially derived by
  comparing the size of $F$ and the parameter $k$. Summarizing all
  these steps gives the proof of the theorem.
\end{proof}
\end{theorem}

Combining Theorem~\ref{new13} and Lemma~\ref{lem-ker-2degsafty}, we
have

\begin{corollary}
\label{chencor22}
There is an $O(n^2 \log^6 n)$-time algorithm that on an instance
$(G; V_1, V_2; k)$ of {\sc disjoint-fvs} where all vertices in $V_1$
have degree bounded by 3, either constructs a $V_1$-FVS of size
bounded by $k$, if such a FVS exists, or reports correctly that no
such a $V_1$-FVS exists.
\end{corollary}

We remark that Corollary~\ref{chencor22} is the best possible in terms
of the maximum vertex degree in the set $V_1$. This can be reasoned as
follows.  It is known that the {\sc fvs} problem on graphs of maximum
degree $4$ is NP-hard \cite{speckenmeyer-83-thesis-fvs}.  Given an
instance $G$ of the {\sc fvs} problem on graphs of maximum degree $4$,
we add a degree-$2$ vertex to the middle of each edge in $G$.  Let the
new graph be $G'$.  Let $V_1$ be the set of vertices in $G'$ that
correspond to the original vertices in $G$, and let $V_2$ be the set
of new degree-$2$ vertices in $G'$.  Now it is rather straightforward
to see that a minimum $V_1$-FVS in $G'$ corresponds to a minimum FVS
in the original graph $G$. Moreover, the maximum vertex degree in the
set $V_1$ in $G'$ is bounded by $4$. This proves that the {\sc
  disjoint-fvs} problem is NP-hard even when restricted to graphs in
which the maximum vertex degree in the set $V_1$ is $4$.

\section{An improved algorithm for {\sc disjoint-fvs}}
\label{sec:unwgraphs}

Now we consider {\sc disjoint-fvs} in general. Let $(G; V_1, V_2; k)$
be an instance of {\sc disjoint-fvs}, for which we are looking for a
$V_1$-FVS of size bounded by $k$. Our algorithm for solving the {\sc
  disjoint-fvs} problem is presented in Figure~\ref{FVS-Algorithm}.

\begin{figure}[h]
\setbox4=\vbox{\hsize28pc \noindent\strut
\begin{quote}
\vspace*{-5mm}
\small 
{\bf Algorithm Feedback($G,V_1,V_2, k$)}\\
{\sc input}: an instance $(G; V_1, V_2; k)$ of {\sc disjoint-fvs}.\\
$\backslash\backslash$ $p =$ the number of nice $V_1$-vertices;
    $\tau_2 =$ the number of components in $G[V_2]$. \\
{\sc output}: a $V_1$-FVS $F$ of size bounded by $k$ in $G$ if such
 a $V_1$-FVS exists, or ``No'' otherwise.

1. \hspace*{3mm} {\bf if} $(k<0)$ or ($k=0$ and $G$ is not a forest)
     or ($2p \geq 2k + \tau_2$)
 {\bf then} return ``No'';\\
2. \hspace*{3mm} {\bf if} ($k\ge 0$ and $G$ is a forest) or ($p=|V_1|$)
   {\bf then} solve the problem in polynomial time;\\
3. \hspace*{3mm} {\bf if} a vertex $w\in V_1$ has degree $\leq 1$
    {\bf then} return {\bf Feedback}$(G-w,V_1 \setminus \{w\}, V_2, k)$; \\
4. \hspace*{3mm} {\bf if} a vertex $w\in V_1$ has two neighbors in the same
      component in $G[V_2]$ \\
   \hspace*{7mm} {\bf then} return $\{w\} \cup \mbox{\bf Feedback}(G-w,V_1 \setminus \{w\}, V_2,k-1)$;\\
5. \hspace*{3mm} {\bf if} a vertex $w \in V_1$ has degree 2 {\bf then}\\
  \hspace*{11mm}  return {\bf Feedback}$(G', V_1, V_2, k)$, where $G' = G$ with
     the vertex $w$ smoothened;\\
6. \hspace*{3mm} {\bf if} a leaf $w$ in $G[V_1]$ is not a nice
     $V_1$-vertex and has $\geq 3$ neighbors in $V_2$ {\bf then} \\
6.1 \hspace*{6mm}    $F_1=$ {\bf Feedback}$(G-w,V_1 \setminus \{w\},V_2,k-1)$; \\
6.2 \hspace*{6mm}   {\bf if} $F_1 \ne$ ``No'' {\bf then} return $F_1\cup\{w\}$\\
6.3 \hspace*{6mm}   {\bf else} return {\bf Feedback}$(G,V_1 \setminus \{w\}, V_2 \cup \{w\},k)$;\\
7. \hspace*{3mm} pick a lowest parent $w$ in any tree in $G[V_1]$ and let $v$ be a child of $w$;\\
7.1 \hspace*{6mm}   $F_1=$ {\bf Feedback}$(G-w,V_1\setminus\{w,v\},V_2\cup\{v\},k-1)$;\\
7.2 \hspace*{7mm}{\bf if} $F_1\ne$ ``No'' {\bf then}  return $F_1\cup\{w\}$\\
7.3 \hspace*{7mm}{\bf else} return
    {\bf Feedback}$(G,V_1\setminus\{w\},V_2\cup\{w\},k)$.
\end{quote} \vspace*{-5mm} \strut} $$\boxit{\box4}$$
\vspace*{-7mm}
 \caption{Algorithm for {\sc disjoint-fvs}}
\label{FVS-Algorithm}
\end{figure}

We first give some explanations to the terminologies used in the
algorithm.  A vertex $v$ in the set $V_1$ is a {\it nice $V_1$-vertex}
if $v$ is of degree $3$ and if all its neighbors are in the set
$V_2$. We will denote by $p$ the number of nice $V_1$-vertices in $G$,
and, as before, by $\tau_2$ the number of components in the
induced subgraph $G[V_2]$. We have slightly abused the use of the set
union operation in step 4 in the sense that when {Feedback$(G-w,
  V_1 \setminus \{w\}, V_2, k-1)$} returns ``No,'' then the union
$\{w\}\cup\mbox{Feedback}(G-w,V_1 \setminus \{w\},V_2,k-1)$ is
also interpreted as a ``No.'' In step 5, by ``smoothening'' a
degree-$2$ vertex $w$, we mean replacing the vertex $w$ and the two
edges incident to $w$ with a new edge connecting the two neighbors of
$w$. In step 6, by a ``leaf'' in $G[V_1]$, we mean a vertex $w$ that
has at most one neighbor in the set $V_1$. Finally, in step 7, we
assume that we have picked an (arbitrary) vertex in each tree in
$G[V_1]$ and designate it as the root of the tree so that a
parent-child relationship is defined in the tree. A ``lowest parent''
$w$ in a tree in $G[V_1]$ is a vertex in the tree that has children
and all its children are leaves.

We start with the following lemma.

\begin{lemma}
\label{boundary} If $2 p \geq 2 k + \tau_2$, then there is no $V_1$-FVS
of size bounded by $k$ in the graph $G$.

\begin{proof}
  Suppose that there is a $V_1$-FVS $F$ of size $k' \leq k$. Let
  $V_1'$ be the set of any $p - k'$ nice $V_1$-vertices that are not
  in $F$.  Then the subgraph $G' = G[V_2 \cup V_1']$ induced by the
  vertex set $V_2 \cup V_1'$ is a forest. On the other hand, the
  subgraph $G'$ can be constructed from the induced subgraph $G[V_2]$
  and the $p-k'$ isolated vertices in $V_1'$, by adding the $3(p-k')$
  edges that are incident to the vertices in $V_1'$. Since $k' \leq
  k$, we have $2(p-k') \geq 2(p-k) \geq \tau_2$. This gives $3(p-k') =
  2(p-k') + (p-k') \geq \tau_2 + (p-k')$. This contradicts the fact
  that $G'$ is a forest---in order to keep $G'$ a forest, we can add
  at most $\tau_2 + (p-k') - 1$ edges to the structure that consists
  of the induced subgraph $G[V_2]$ of $\tau_2$ components
  and the $p-k'$ isolated vertices in $V_1'$.  This contradiction
  proves the lemma.
\end{proof}
\end{lemma}

Now we are ready to analyze the algorithm {Feedback}$(G, V_1, V_2, k)$
for the {\sc disjoint-fvs} problem in Figure~\ref{FVS-Algorithm}. We first
prove the correctness of the algorithm.

\begin{lemma}
\label{correctness}
The algorithm {Feedback} solves the {\sc disjoint-fvs} problem correctly.

\begin{proof}
  The correctness of step 1 follows from Lemma~\ref{boundary} and
  other trivial facts. If $k \geq 0$ and the graph $G$ is a forest,
  then obviously the empty set $\emptyset$ is a solution to the input
  instance. If $p = |V_1|$, then by definition, all vertices in the
  set $V_1$ have degree $3$. By Corollary~\ref{chencor22}, this case
  can be solved in polynomial time. This verifies the correctness of
  step 2.  The correctness of step 3 follows from the fact that no
  vertices of degree bounded by $1$ can be contained in any
  cycle. Step 4 is correct because in this case, the vertex $w$ is the
  only vertex in the set $V_1$ in a cycle in the graph $G$, so it must
  be included in the objective $V_1$-FVS.  Step 5 follows from
  Lemma~\ref{lem-ker-2degsafty} and the fact that step 4 does apply to
  the vertex $w$.

  Step 6 is correct because it simply branches on either including or
  excluding the vertex $w$ in the objective $V_1$-FVS. Note that after
  passing steps 3-5, all vertices in the set $V_1$ have degree at
  least $3$, and after passing steps 3-6, each vertex in the set $V_1$
  either is a nice $V_1$-vertex or has at least one neighbor in
  $V_1$. In particular, after steps 3-6, if a leaf $v$ in $G[V_1]$ is
  not a nice $V_1$-vertex, then $v$ has exactly two neighbors in $V_2$
  that belong to two different components of $G[V_2]$. Now consider
  step 7. As remarked above (also noting step 2), at this point there
  must be a tree with more than one vertex in the induced subgraph
  $G[V_1]$. Therefore, we can always find a lowest parent $w$ in a
  tree in $G[V_1]$. Step 7 branches on this lowest parent $w$. In case
  $w$ is included in the objective $V_1$-FVS, $w$ is deleted from the
  graph, and the parameter $k$ is decreased by $1$.  Note that after
  the vertex $w$ is deleted, the child $v$ of $w$ becomes of degree
  $2$ with its two neighbors in two different components of
  $G[V_2]$. By Lemma~\ref{JChen}, the vertex $v$ can be
  excluded from the objective $V_1$-FVS. Thus, it is safe to move the
  vertex $v$ from set $V_1$ to set $V_2$. This verifies the
  correctness of steps 7.1-7.2.  Step 7.3 is simply to exclude the
  vertex $w$ from the objective $V_1$-FVS.

Observe that before making recursive calls, each of the steps 3-7
decreases the number of vertices in the set $V_1$ by at least $1$. Therefore,
the algorithm must terminate in a finite number of steps. Summarizing all
the above discussion, we conclude with the correctness of the algorithm
{Feedback$(G, V_1, V_2, k)$}.
\end{proof}
\end{lemma}

Now we analyze the complexity of the algorithm {Feedback}. The
recursive execution of the algorithm can be depicted as a search tree
$\cal T$, whose complexity can be analyzed by counting the number of
leaves in the search tree. For an input instance $(G, V_1, V_2, k)$,
we, as before, let $p$ be the number of nice $V_1$-vertices in $G$,
and let $\tau_2$ be the number of components in the induced
subgraph $G[V_2]$. To analyze the complexity of the algorithm more
precisely, we introduce a new measure, defined as $\mu =
2(k-p)+\tau_2$. Let $T(\mu)$ be the number of leaves in the search
tree $\cal T$ for the algorithm on the input $(G, V_1, V_2, k)$.

\begin{theorem}
\label{FVS2} The algorithm {Feedback}$(G, V_1, V_2, k)$ correctly
solves the {\sc disjoint-fvs} problem in time $O(2^{k+ \tau_2/2}n^2 \log^6 n)$,
where $n$ is the number of vertices in the graph $G$, and $\tau_2$ is
the number of components in the induced subgraph $G[V_2]$.

\begin{proof}
  We have verified the correctness of the algorithm in
  Lemma~\ref{correctness}.  Herein we analyze its complexity, i.e., we
  consider the value $T(\mu)$.

  Each of steps 1-5 of the algorithm proceeds without branching; hence
  it suffices to verify that neither of them {\it increases} the value
  of the measure $\mu$.  Step 3 does not change the values of $k$,
  $p$, and $\tau_2$, thus neither that of $\mu$. Step 4 does not
  change the value $\tau_2$, but decreases the value $k$ by
  $1$. Moreover, step 4 {\it may} also decrease the value $p$ by at
  most $1$ (in case the vertex $w$ is a nice $V_1$-vertex). Overall,
  step 4 does not increase the value $\mu = 2(k-p) + \tau_2$. Step 5
  does not change the value of $k$. Moreover, it will never decrease
  the value of $p$ or increase the value of $\tau_2$. Note that step 5
  may increase the value of $p$ (e.g., a neighbor of $w$ in $V_1$ may
  become a nice $V_1$-vertex after smoothening $w$) or decrease the
  value of $\tau_2$ (e.g., when the two neighbors of $w$ are in two
  different components in $G[V_2]$).  In any case, step 5
  does not increase the value $\mu = 2(k-p) + \tau_2$.

Now we study the branching steps. First consider step 6. The branch of
steps 6.1-6.2 decreases the value $k$ by $1$ and does not change the
value of $\tau_2$. Moreover, the steps may increase the value of $p$
(e.g., a neighbor of $w$ in $V_1$ may become a nice $V_1$-vertex after
deleting $w$ from the graph) but will never decrease
the value of $p$. Therefore, the branch of steps 6.1-6.2 will decrease
the value $\mu = 2(k-p) + \tau_2$ by at least $2$. On the other hand,
because $w$ has at least three neighbors in $V_2$, step 6.3 will decrease
the value of $\tau_2$ by at least $2$, while neither changing the value
of $k$ nor decreasing the value of $p$. Thus, step 6.3 also decreases
the value $\mu = 2(k-p) + \tau_2$ by at least $2$. In summary, if
step 6 is executed in the algorithm, then the function $T(\mu)$ satisfies
the recurrence relation $T(\mu) \leq 2 T(\mu - 2)$.

Similarly, the branch of steps 7.1-7.2 deletes the vertex $w$ from the
graph and decreases the value of $k$ by $1$. As we pointed out before,
since the algorithm has passed steps 3-6, the leaf $v$ has exactly three
neighbors: one is $w$ and the other two are in two different
components in $G[V_2]$. Therefore, after deleting $w$ from
the graph, moving the degree-$2$ vertex $v$ from set $V_1$ to set $V_2$
decreases the value of $\tau_2$ by $1$. Also note that in this branch,
the value of $p$ is not changed (because of step 6, the vertex $w$ cannot
have a neighbor that is a leaf in $G[V_1]$ but has three neighbors in
$V_2$).  In summary, the branch of steps 7.1-7.2 decreases the value
$\mu = 2(k-p) + \tau_2$ by at least $3$. Now consider step 7.3 that moves
the vertex $w$ from set $V_1$ to set $V_2$. We break this case into two
subcases:

Subcase 7.3.1. The vertex $w$ has at least one neighbor in $V_2$. Then
moving $w$ from $V_1$ to $V_2$ neither changes the value of $k$ nor
increases the value of $\tau_2$. On the other hand, it creates at least
one new nice $V_1$-vertex (i.e., the vertex $v$) thus increases the value
of $p$ by at least $1$. Therefore, in this subcase, step 7.3 decreases
the value of $\mu = 2(k-p) + \tau_2$ by at least $2$.

Subcase 7.3.2. The vertex $w$ has no neighbor in $V_2$. Because the
degree of $w$ is larger than $2$ and $w$ is a lowest parent in
$G[V_1]$, $w$ has at least two children in $V_1$, each is a leaf in
$G[V_1]$ with exactly two neighbors that are in two different
components of $G[V_2]$.  Note that after moving $w$ from $V_1$ to
$V_2$, all children of $w$ in $G[V_1]$ will become nice
$V_1$-vertices. Therefore, moving $w$ from $V_1$ to $V_2$ increases
the value of $\tau_2$ by $1$, and increases the value of $p$ by at
least $2$, with the value of $k$ unchanged. Therefore, in this
subcase, step 7.3 decreases the value of $\mu = 2(k-p) + \tau_2$ by at
least $3$.

Summarizing the above discussion, we conclude that if step 7 is executed
in the algorithm, then the function $T(\mu)$ satisfies the recurrence
relation $T(\mu) \leq T(\mu - 2) + T(\mu - 3)$.

Therefore, the function $T(\mu)$, which is the number of leaves in the
search tree $\cal T$, in the worst case satisfies the recurrence
relation $T(\mu) \leq 2 T(\mu - 2)$. Also note that
Lemma~\ref{boundary}, if $\mu = 2(k-p) + \tau_2 \leq 0$, then we can
conclude immediately without branching that the input instance is a
``No.'' Therefore, $T(\mu) = 1$ for $\mu \leq 0$. Now the recurrence
relation $T(\mu) \leq 2 T(\mu - 2)$ with $T(\mu) = 1$ for $\mu \leq 0$
can be solved using the well-known techniques in parameterized
computation (see, for example, \cite{downey-fellows-13}), as
follows. The characteristic polynomial for the recurrence relation
$T(\mu) = 2 T(\mu - 2)$ is $x^2 - 2$, which has a unique positive root
$\sqrt{2}$. From this, we derive $T(\mu) = (\sqrt{2})^{\mu} =
2^{\mu/2}$.  Moreover, it is fairly easy to see that each
computational path in the search tree $\cal T$ has its time bounded by
$O(n^2 \log^6 n)$, and $\mu/2 = k - p + \tau_2/2 \leq k +
\tau_2/2$. Therefore, the running time of the algorithm {\bf
  Feedback}$(G, V_1, V_2, k)$ is $O(2^{k+ \tau_2/2}n^2 \log^6 n)$
\end{proof}
\end{theorem}

\section{An improved algorithm for {\sc fvs}}

The results in previous sections lead to an improved algorithm for the
general {\sc fvs} problem. Following the idea of {\it iterative
  compression} proposed by Reed et
al.~\cite{reed-04-odd-cycle-transversals}, we formulate the following
problem:
\begin{quote}
{\sc fvs reduction:} given a graph $G$ and a FVS $F$ of size $k+1$
for $G$, either construct a FVS of size bounded by $k$ for $G$, or
report that no such a FVS exists.
\end{quote}

\begin{lemma}
\label{FVS3} The {\sc fvs reduction} problem can be solved in time $O^*(3.83^k)$.

\begin{proof}
The proof goes similar to that for Lemma 2 in [3]. Let $G = (V, E)$ be
a graph and let $F_{k+1}$ be a FVS of size $k+1$ in $G$. Suppose that
the graph $G$ has a FVS $F_k'$ of size $k$, and let the intersection
$F_{k+1} \cap F_k'$ be a set $F_{k-j}$ of $k-j$ vertices, for some $j$,
$0 \leq j \leq k$. Let $F_{j+1} = F_{k+1} \setminus F_{k-j}$ and
$F_j' = F_k' \setminus F_{k-j}$. Construct the graph $G' = G - F_{k-j}$.
Note that both $F_{j+1}$ and $F_j'$ are FVS for $G'$, and that $F_{j+1}$
and $F_j'$ are disjoint. Thus, if we let $V_1' = V \setminus F_{k+1}$
and $V_2' = F_{j+1}$, then $F_j'$ is a solution to the instance
$(G', V_1', V_2', j)$ of the {\sc disjoint-fvs} problem. On the other
hand, it is also easy to see that any solution to the instance
$(G', V_1', V_2', j)$ of {\sc disjoint-fvs} plus the subset $F_{k-j}$ makes
a FVS of no more than $k$ vertices for the original graph $G$.

Therefore, to solve the instance $(G, F_{k+1})$ for the {\sc fvs reduction}
problem, it suffices to find the subset $F_{k-j} = F_{k+1} \cap F_k'$ of $k-j$ vertices in
$F_{k+1}$ for some integer $j$, $0 \leq j \leq k$, then to solve the
instance $(G', V_1', V_2', j)$ for the {\sc disjoint-fvs} problem. To
find the subset $F_{k-j}$ of $F_{k+1}$, we enumerate all subsets of $k-j$
vertices in $F_{k+1}$ for all $0 \leq j \leq k$. To solve the corresponding
instance $(G', V_1', V_2', j)$ for {\sc disjoint-fvs} derived from the
subset $F_{k-j}$ of $F_{k+1}$, we call the algorithm
{Feedback$(G', V_1', V_2', j)$}. By Theorem~\ref{FVS2} (note that
$\tau_2 \leq |V_2'| = j+1$), the instance $(G', V_1', V_2', j)$ for
{\sc disjoint-fvs} can be solved in time $O(2^{j + (j+1)/2}n^2 \log^6 n) =
O(2.83^j n^2 \log^6 n)$. Applying this procedure for every integer $j$
($0 \leq j \leq k$) and all subsets of size $k-j$ in $F_{k+1}$ will
successfully find a FVS of size $k$ in the graph $G$, if such a
FVS exists. This algorithm solves the {\sc fvs reduction} problem in time
$\sum_{j=0}^k {{k+1} \choose {k-j}} \cdot O(2.83^j n^2 \log^6 n) =
O^*(3.83^k)$.
\end{proof}
\end{lemma}

Finally, by combining Lemma~\ref{FVS3} with the iterative compression
technique \cite{reed-04-odd-cycle-transversals,chen-08-ufvs}, we
obtain the main result of this paper, which solves the {\sc fvs}
problem, formally defined as follows:

\begin{quote}
 {\sc fvs:} given a graph $G$ and a parameter $k$, either construct
   a FVS of size bounded by $k$ for the graph $G$, or report that no
   such FVS exists.
\end{quote}

\begin{theorem}\label{thm:mainunweighted}
The {\sc fvs} problem is solvable in time $O^*(3.83^k)$.

\begin{proof}
To determine if a given graph $G = (V, E)$ has a FVS of size bounded
by $k$, we start by applying the polynomial-time approximation algorithm of approximation
ratio $2$ for the {\sc minimum feedback vertex set} problem \cite{bafna-99-approximate-fvs}.
This algorithm runs in $O(n^2)$ time, and either returns a FVS $F'$ of
size at most $2k$, or verifies that no FVS of size bounded by $k$ exists.
Thus, if no FVS is returned by the algorithm, then no FVS of size bounded
by $k$ exists. In the case of the opposite result, we use any subset $V'$
of $k$ vertices in $F'$, and put $V_0 = V' \cup (V \sm F')$. Obviously,
the induced subgraph $G[V_0]$ has a FVS $V'$ of size $k$. Let
$F' \sm V' = \{v_1, v_2, \ldots, v_{|F'|-k}\}$, and let
$V_i = V_0 \cup \{v_1, \ldots, v_i\}$ for $i \in \{0, 1, \ldots, |F'|-k\}$.
Inductively, suppose that we have constructed a FVS $F_i$ for the graph
$G[V_i]$, where $|F_i| = k$. Then the set $F_{i+1}' = F_i \cup\{v_{i+1}\}$
is a FVS for the graph $G[V_{i+1}]$, and $|F_{i+1}'| = k+1$.

Now the pair $(G[V_{i+1}], F_{i+1}')$ is an instance for the {\sc
fvs reduction} problem. Therefore, in time $O^*(3.83^k)$, we can
either construct a FVS $F_{i+1}$ of size $k$ for the graph
$G[V_{i+1}]$, or report that no such a FVS exists. Note that if the
graph $G[V_{i+1}]$ does not have a FVS of size $k$, then the
original graph $G$ cannot have a FVS of size $k$. In this case, we
simply stop and claim the non-existence of a FVS of size $k$ for
the original graph $G$. On the other hand, with a FVS $F_{i+1}$ of
size $k$ for the graph $G[V_{i+1}]$, our induction proceeds to the
next graph $G[V_{i+1}]$, until we reach the graph $G =
G[V_{|F'|-k}]$. This process runs in time $k \cdot O^*(3.83^k)
= O^*(3.83^k)$ since $|F'|-k \leq k$, and solves the {\sc fvs} problem.
\end{proof}
\end{theorem}

\section{Concluding remarks}
We developed an $O^*(3.83^{k})$-time parameterized algorithm for the
{\sc fvs} problem.  Our algorithm was obtained by a nontrivial
combination of several known techniques in algorithm research and
their generalizations.  This includes iterative compression,
branch-and-search, and efficient algorithms for graphs of low
vertex-degrees. For branch-and-search processes for dealing with the
{\sc fvs} problem, we introduced new branching rules and new branching
measures, which allow us to more effectively reduce a general instance
into a polynomial-time solvable instance of the problem and to more
accurately evaluate the efficiency of the branch-and-search process.
For efficient algorithms for graphs of low vertex-degrees, we use a
nontrivial reduction that transforms the {\sc fvs} problem to a
polynomial-time solvable version of the matroid matching problem. Note
that using matroid matching to solve the {\sc fvs} problem for
$3$-regular graphs has been observed previously
\cite{speckenmeyer-88-fvs-and-nsis,ueno-88-cubic-fvs,furst88}, while
we extended the techniques to solve the {\sc disjoint-fvs} problem on
a larger graph class in which not all vertices are required to have
degree bounded by $3$.

Further faster algorithms for {\sc fvs} have drawn much attention in
the recent research in parameterized
computation~\cite{dagstuhl2013}. Following our approach with a new
reduction rule introduced, Kociumaka and Pilipczuk \cite{newfvs} have
announced a revision of our algorithm that has an improved running
time $O^*(3.62^k)$ for the {\sc fvs} problem. On the other hand, the
study on the lower bound of the {\sc fvs} problem has made significant
progress. Based on the {\em Strong Exponential Time Hypothesis} (see
\cite{lokshtanov-11-lower-bound-on-ETH}), Cygan et
al.~\cite{cygan-11-connectivity-treewidth} have reported a lower bound
on the complexity of the {\sc fvs} problem in terms of the pathwidth
$pw$ of a graph, which states that the {\sc fvs} problem cannot be
solved in time $O^*((3-\epsilon)^{pw})$ for any positive constant
$\epsilon > 0$. This result does not yet directly lead to a lower
bound for the {\sc fvs} problem in terms of the parameter $k$, which
is the {\it size} of the objective FVS (to see this, observe that the
ladder graph $P_l \times P_2$ has a pathwidth $2$ but its minimum FVS
has a size $\lfloor l/2 \rfloor$, where $P_i$ denotes the simple path
of $i$ vertices). On the other hand, studying the complexity of the
{\sc fvs} problem in terms of graph pathwidth or treewidth seems to
have very interesting connection to the complexity of the original
{\sc fvs} problem.  For example, the $O^*(3^{tw})$-time randomized
algorithm for the {\sc fvs} problem proposed
in~\cite{cygan-11-connectivity-treewidth}, where $tw$ is the treewidth
of the input graph, directly implies an $O^*(3^k)$-time randomized
algorithm for {\sc fvs}.  In particular, this has motivated an
interesting open problem whether there is a deterministic
$O^*(3^k)$-time algorithm for the {\sc fvs}
problem~\cite{dagstuhl2013}.

It is interesting to observe that the research on parameterized
algorithms and that on approximation algorithms for the {\sc fvs}
problem have undergone a similar process. Early algorithms used the
cycle packing-covering duality, and hence ended with
$O^*(\log{k}^{O(k)})$-time parameterized algorithms
\cite{raman-02-fvs,kanj-04-fvs} and $O(\log{n})$-ratio approximation
algorithms \cite{erdos-62-number-disjoint-circuits},
respectively. Later algorithms turned to the observation on graph
vertex-degrees, which resulted in $O^*(2^{O(k)})$-time parameterized
algorithms \cite{chen-08-ufvs,dehne-05-fvs} and constant-ratio
approximation algorithms
\cite{bafna-99-approximate-fvs,becker-96-approximate-fvs},
respectively. However, constant-ratio approximation algorithms for
{\sc fvs} do not seem to rely on a process that is related to the
iterative compression process \cite{reed-04-odd-cycle-transversals},
which, on the other hand, seems to have played a critical role in the
development of all $O^*(2^{O(k)})$-time parameterized algorithms for
the {\sc fvs} problem. A parameterized algorithm based on iterative
compression for the {\sc fvs} problem runs in time
$O^*((1+\alpha)^k)$, where $\alpha$ is a constant such that the {\sc
  disjoint-fvs} problem can be solved in time $O^*(\alpha^k)$.  Since
the {\sc disjoint-fvs} problem is NP-hard, the constant $\alpha$ has
to be larger than $1$.  In other words, using the iterative
compression technique excludes the possibility of solving the {\sc
  fvs} problem in time $O^*(2^k)$.  An interesting research direction
and a possible approach to developing further improved algorithms for
the {\sc fvs} problem is to explore new algorithmic techniques that
are {\it not} based on iterative compression.

\paragraph{Acknowledgment.}  
We would like to thank anonymous referees for thoughtful and detailed
comments, which led to an improved presentation.

\end{document}